\documentclass[prd,preprint,tightenlines,floatfix,
preprintnumbers,nofootinbib,eqsecnum,superscriptaddress]{revtex4-1}

 \usepackage[dvips,final]{graphicx}
  \usepackage{amssymb}
   \usepackage{amsmath}
    \usepackage{amsfonts}
     \usepackage{epsfig}
      \usepackage{bm}

\usepackage{soul}

\usepackage[section]{placeins}

\usepackage{multirow}
\usepackage{ctable}
\usepackage{booktabs}
\usepackage{array}
\usepackage{tabularx}

\newcommand{\bp}{\mbox{\boldmath $p$}}
\newcommand{\bb}{\mbox{\boldmath $b$}}
\newcommand{\bq}{\mbox{\boldmath $q$}}

\newcommand{\bP}{\mbox{\boldmath $P$}}
\newcommand{\bQ}{\mbox{\boldmath $Q$}}
\newcommand{\bB}{\mbox{\boldmath $B$}}
\newcommand{\bR}{\mbox{\boldmath $R$}}

\newcommand{\bs}{\mbox{\boldmath $s$}}

\newcommand{\bE}{\mbox{\boldmath $E$}}

\newcommand{\half}{{1\over 2}}

\begin{document}

\title{Centrality dependence of dilepton production via $\gamma \gamma$ processes  from
Wigner distributions of photons in nuclei 
}

\author{Mariola K{\l}usek-Gawenda}
\email{Mariola.Klusek@ifj.edu.pl}
\affiliation{Institute of Nuclear Physics Polish Academy of Sciences,
ul. Radzikowskiego 152,
PL-31-342 Krak\'ow, Poland}

\author{Wolfgang Sch\"afer }
\email{Wolfgang.Schafer@ifj.edu.pl}
\affiliation{Institute of Nuclear Physics Polish Academy of Sciences,
ul. Radzikowskiego 152, PL-31-342 Krak\'ow, Poland}

\author{Antoni Szczurek}
\email{Antoni.Szczurek@ifj.edu.pl}
\affiliation{Institute of Nuclear Physics Polish Academy of Sciences, ul. Radzikowskiego 152,
PL-31-342 Krak\'ow, Poland}
\affiliation{
College of Natural Sciences, Institute of Physics,
University of Rzesz\'ow, ul. Pigonia 1, PL-35-310 Rzesz\'ow, Poland}

\begin{abstract}
We propose a new complete method, based on the Wigner
distributions of photons, how to calculate differential distributions of
dileptons created via photon-photon fusion in semicentral ($b<2R_A$)
$AA$ collisions. 
The formalism is used to calculate different distributions of invariant mass, dilepton transverse momentum and acoplanarity for different
regions of centrality. The results of calculation are compared
with recent STAR, ALICE and ATLAS experimental data.
Very good agreement with the data is achieved
without free parameters and
without including additional mechanisms such as a possible rescattering of leptons in the quark-gluon plasma. 
\end{abstract} 

\maketitle

\section{Introduction}

Ultrarelativistic Heavy Ions of large charge $Z$ are accompanied by a large flux of Weizs\"acker--Williams photons.
This opens up the opportunity to study a variety of photoinduced nuclear processes, as well as photon-photon processes. See for example the reviews with focus on RHIC and LHC \cite{Baur:2001jj,Bertulani:2005ru,Contreras:2015dqa,Klein:2020fmr,Schafer:2020bnm}. 
Until recently, most investigations have focused on
ultraperipheral collisions, 
where the coherent enhancement $\propto Z^2$ of Weizs\"acker-Williams fluxes is evident. 
Here the impact parameter of the collision satisfies $b > 2 R_A$, with $R_A$ being the nuclear radius. A prominent role play the dileptons created via photon-photon fusion, for recent calculations at collider kinematics, see for example 
\cite{KlusekGawenda:2010kx,Azevedo:2019fyz}.

However, if one interprets Weizs\"acker-Williams photons as partons of the nucleus, it appears natural that the coherent photon cloud also contributes in semicentral or central collisions, where $b<2 R_A$. In such collisions the colliding nuclei will interact strongly generating an "underlying event" for the 
$\gamma \gamma$ process, which may include the production of a quark-gluon plasma (QGP). In fact, such contributions have long been suggested \cite{Baron:1993sk}. First experimental evidence for the relevance of photoproduction processes in inelastic Heavy Ion reactions
was reported by the ALICE collaboration \cite{Adam:2015gba}. Indeed, the large enhancement of $J/\psi$ at low transverse momenta observed in \cite{Adam:2015gba} can be readily explained through the presence of 
photoproduction mechanisms \cite{Klusek-Gawenda:2015hja}.

It was realized only recently that also coherent photon-photon processes survive 
in semicentral collisions. Two years ago the STAR
collaboration at RHIC \cite{Adam:2018tdm} observed a large enhancement at very low transverse momenta
of the dielectron pair, $P_T < 150 \, \rm{MeV}$ . This enhancement was interpreted soon as due to photon-photon fusion \cite{Klusek-Gawenda:2018zfz,Klein:2018fmp,Zha:2018tlq,Li:2019sin}.
Dilepton production in heavy ion collisions is traditionally considered as a source of information on the properties of the QCD matter produced in the collision \cite{Tserruya:2009zt,Rapp:2013nxa}.
Besides the contributions of medium-modified vector mesons and thermal radiation from the QGP, 
there are several other mechanisms of dilepton production in
ultrarelativistic heavy ion collisions.
These are conventionally subsumed as the hadronic cocktail contribution which involves essentially
all sources in nucleon-nucleon collisions (Dalitz decays, vector meson 
production, Drell-Yan mechanism, semileptonic decays of pairs of mesons).

In \cite{Klusek-Gawenda:2018zfz} we performed a comprehensive study of the interplay of all these mechanisms, highlighting the dominance of $\gamma \gamma$ processes at very low $P_T$. 
However, in Ref.\cite{Klusek-Gawenda:2018zfz}, the $P_T$ distribution of di-electrons was obtained using a $k_T$-factorization method.
In such an approach
the distribution in $P_T$ is independent of the impact parameter
(and therefore centrality), and only the normalization contains information on the latter.

In \cite{Zha:2018tlq,Li:2019sin} it has been proposed to use an approach that has been put forward long ago in \cite{Vidovic:1992ik}. However the formalism presented in \cite{Vidovic:1992ik} applies only to the impact parameter dependence of invariant mass distributions of dileptons, and cannot be readily employed in an "unintegrated" form to yield transverse momentum distributions.

In \cite{Klein:2018fmp} it was suggested that rescattering in QGP may lead
to a broadening of the $P_T$ peak and acoplanarity distribution of dileptons.

In this paper \footnote{Preliminary results of this work have been presented in [WS - 40th International Conference on High Energy Physics ICHEP2020, 30.07-5.08.2020 virtual meeting, Prague, MKG - Zimányi School Winter Workshop 2020, 7-10.12.2020 virtual meeting Budapest]}  we present a more complete approach that allows to calculate the centrality dependence of the $\gamma \gamma$ mechanism of dilepton production. It is based on so-called
Wigner distributions \footnote{a similar approach based on Wigner functions has recently been obtained in \cite{Klein:2020jom}.}. We present the formalism in the next section. Then we
shall confront results of the new method with existing experimental data. Conclusions will close our paper.

\section{Centrality dependent cross section of dilepton production}

\begin{figure}[!h]
    \centering
    \includegraphics[width=.5\textwidth]{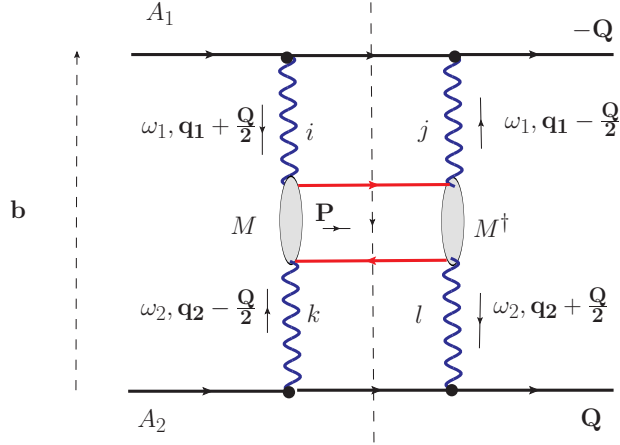}
    \caption{The cut off-forward $A_1 A_2 \to A_1 A_2$ amplitude. Its Fourier transform w.r.t. $\bQ$ yields the impact-parameter dependent cross section of dilepton production.}
    \label{fig:Feynman}
\end{figure}
For the ion moving at longitudinal boost $\gamma$, let us denote the electric field vector:
\begin{eqnarray}
\bE (\omega,\bq) = Z \sqrt{\frac{\alpha_{em}}{\pi}} \,  {\bq F_{\rm ch}(\bq^2+q^2_\parallel) \over \bq^2 + q^2_\parallel} \, , {\rm where} \, q_\parallel = {\omega \over \gamma} \, ,
\end{eqnarray}
and $F_{\rm ch}(Q^2)$ is the charge form factor of the nucleus.

The impact parameter dependent cross section of interest is obtained from the Fourier transform of the cut off-forward 
$A_1 A_2 \to A_1 A_2$ amplitude shown in Fig.\ref{fig:Feynman}.

The relevant factorization formula can be written in terms of the Wigner function
\begin{eqnarray}
N_{ij} (\omega,\bb,\bq) &=& \int {d^2 \bQ \over (2 \pi)^2} \, \exp[-i \bb \bQ] 
\, E_i \Big(\omega, \bq + {\bQ \over 2} \Big) E^*_j \Big(\omega, \bq - {\bQ \over 2} \Big) \, 
\label{eq:Wigner}
\\
&=& \int d^2\bs \,  \exp[i \bq \bs] \,  E_i \Big(\omega, \bb + {\bs \over 2} \Big) E^*_j \Big(\omega, \bb - {\bs \over 2} \Big) .
\label{eq:flux}
\end{eqnarray}
It has the property of being a density matrix in photon polarizations (here we use the basis of cartesian linear polarizations), and depends on impact parameter and transverse momentum.

Being a Wigner function, the standard photon fluxes in momentum space and impact parameter space, are obtained after 
integration over impact parameter or momentum space, respectively, a summation over photon polarizations is implied:
\begin{eqnarray}
N(\omega,\bq) &=& \delta_{ij} \int d^2\bb \, N_{ij} (\omega,\bb,\bq) = \delta_{ij} \, E_i (\omega, \bq) E^*_j (\omega, \bq) = \Big| \bE (\omega,\bq) \Big|^2
\, , \nonumber \\
N(\omega,\bb) &=& \delta_{ij} \int {d^2 \bq \over (2 \pi)^2} N_{ij}(\omega,\bb,\bq) = \delta_{ij} \, E_i (\omega, \bb) E^*_j (\omega, \bb) =\Big| \bE (\omega,\bb) \Big|^2 \, .
\end{eqnarray}
Then the cross section for lepton pair production can be written as a convolution
over impact parameters and transverse momenta
\begin{eqnarray}
{d\sigma \over d^2\bb d^2\bP} &=& \int d^2\bb_1 d^2\bb_2 \, \delta^{(2)}(\bb-\bb_1 + \bb_2)  
\int {d^2\bq_1 \over \pi} {d^2\bq_2 \over \pi} \, \delta^{(2)}(\bP-\bq_1 - \bq_2) \nonumber \\
&\times&
\int {d \omega_1 \over \omega_1} {d \omega_2 \over \omega_2} N_{ij} (\omega_1,\bb_1,\bq_1) N_{kl} (\omega_2,\bb_2,\bq_2)  \, \, {1 \over 2 \hat s} \sum_{\lambda \bar \lambda} M^{\lambda \bar \lambda}_{ik} M^{\lambda \bar \lambda \dagger}_{jl} \, d\Phi(l^+ l^-) .
\end{eqnarray}
Here again we sum over cartesian photon polarizations $i,j,k,l$, and the invariant phase space of the leptons is:
\begin{eqnarray}
d\Phi_{l^+ l^-} = (2 \pi)^4 \, \delta^{(4)}(P - p_1 - p_2) \, {d^4 p_1 \over (2 \pi)^3} \delta(p_1^2 - m_l^2)
{d^4 p_2 \over (2 \pi)^3} \delta(p_2^2 - m_l^2) \, .
\end{eqnarray}
We now parametrize
\begin{eqnarray}
\bb_1 = \bR + {\bB \over 2} , \, \, \bb_2 = \bR - {\bB \over 2} \, \Rightarrow d^2\bb_1 d^2\bb_2 = d^2\bR \, d^2\bB.
\end{eqnarray}
Below, we also use the notation $P_T \equiv |\bP|$ for the transverse momentum of the dilepton pair.
Inserting the representation of the generalized Wigner function given in Eq.~(\ref{eq:Wigner}) ,
and integrating out $\bR$ (the delta function puts $\bb = \bB$), we obtain
\begin{eqnarray}
{d\sigma \over d^2\bb d^2\bP} &=& \int {d^2 \bQ \over (2 \pi)^2}  \exp[-i \bb \bQ] \int {d^2\bq_1 \over \pi} {d^2\bq_2 \over \pi} \, \delta^{(2)}(\bP-\bq_1 - \bq_2) \int {d \omega_1 \over \omega_1} {d \omega_2 \over \omega_2}
\nonumber \\
&\times&   E_i \Big(\omega_1, \bq_1 + {\bQ \over 2} \Big) E^*_j \Big(\omega_1, \bq_1 - {\bQ \over 2} \Big)
E_k \Big(\omega_2, \bq_2 - {\bQ \over 2} \Big) E^*_l \Big(\omega_2, \bq_2 + {\bQ \over 2} \Big) \nonumber \\
&\times&   {1 \over 2 \hat s}  \sum_{\lambda \bar \lambda} M^{\lambda \bar \lambda}_{ik} M^{\lambda \bar \lambda \dagger}_{jl} \, d\Phi(l^+ l^-) \, .
\end{eqnarray}
Notice, that our approach predicts ``flow-like'' correlations between the dilepton transverse momentum and the impact parameter $\bb$. 
We defer the discussion of these correlations to a future publication, and in this work we will average over directions of $\bb$.
The cross section in a certain centrality class is obtained by integrating over the corresponding range of impact parameters:
\begin{eqnarray}
d \sigma [{\cal C}] = \int_{b_{\rm min}}^{b_{\rm max}} db \, {d \sigma \over db} \, .
\end{eqnarray}
If we are interested in the cross section in a certain centrality class ${\cal C}$, we can get rid of the integration over $d^2\bb$. We start from integrating over all possible orientations of $\bb$.  
\begin{eqnarray}
\int d^2\bb  \exp[-i \bb \bQ] 
(\dots) \rightarrow 2\pi
\int db b J_0 (b Q) (\dots) \, .
\end{eqnarray}
Integrating over a range $[b_{\rm min}, b_{\rm max}]$ of impact parameters, we obtain
\begin{eqnarray}
\int_{b_{\rm min}}^{b_{\rm max}}  db \, b J_0(b Q) = {1 \over Q^2} \Big( Qb_{\rm max} J_1(Q b_{\rm max}) - Qb_{\rm min}  J_1(Q b_{\rm min}) \Big) \equiv w(Q; b_{\rm max}, b_{\rm min}) \, .
\end{eqnarray}
Thus the cross section for a certain centrality class $\cal C$ is:
\begin{eqnarray}
{d \sigma [{\cal C}]\over d^2\bP} &=& 
\int {d^2 \bQ \over 2 \pi}  w(Q; b_{\rm max}, b_{\rm min}) 
 \int {d^2\bq_1 \over \pi} {d^2\bq_2 \over \pi}\, \delta^{(2)}(\bP-\bq_1 - \bq_2) \int {d \omega_1 \over \omega_1} {d \omega_2 \over \omega_2}
\nonumber \\
&\times&   E_i \Big(\omega_1, \bq_1 + {\bQ \over 2} \Big) E^*_j \Big(\omega_1, \bq_1 - {\bQ \over 2} \Big)
E_k \Big(\omega_2, \bq_2 - {\bQ \over 2} \Big) E^*_l \Big(\omega_2, \bq_2 + {\bQ \over 2} \Big) \nonumber \\
&\times& 
{1 \over 2 \hat s}  \sum_{\lambda \bar \lambda} M^{\lambda \bar \lambda}_{ik} M^{\lambda \bar \lambda \dagger}_{jl} \, d\Phi(l^+ l^-) \, .
\end{eqnarray}
The impact parameter intervals $[b_{\rm min}, b_{\rm max}]$ corresponding to a given centrality class are obtained from a simple optical Glauber approach, as described in more detail in \cite{Klusek-Gawenda:2018zfz}.
Now, we come to the helicity structure of the $\gamma \gamma \to l^+ l^-$ amplitude.
Here, indices $i,j$ correspond to linear polarizations of photons, while $\lambda, \bar \lambda$ are the helicities of leptons.
The amplitude takes the form (below $M(\lambda_1 \lambda_2 , \lambda \bar \lambda) $is the helicity amplitude for the $\gamma(\lambda_1) \gamma(\lambda_2) \to l^-(\lambda) l^+(\bar \lambda)$ process):
\begin{eqnarray}
M^{\lambda \bar \lambda}_{ik} &=& - \half \Big( \hat x_i \hat x_k + \hat y_i \hat y_k \Big)
\Big( M(++, \lambda \bar \lambda) + M(--, \lambda \bar \lambda) \Big) \nonumber \\
&& - {i \over 2} 
\Big( \hat x_i \hat y_k - \hat y_i \hat x_k \Big)
\Big( M(++, \lambda \bar \lambda) - M(--, \lambda \bar \lambda)\Big)
\nonumber \\
&& +{1 \over 2} \Big( \hat x_i \hat x_k - \hat y_i \hat y_k \Big) \Big( M(-+, \lambda \bar \lambda) + M(+-, \lambda \bar \lambda)\Big)
\nonumber \\
&& + {i \over 2} \Big( \hat x_i \hat y_k + \hat y_i \hat x_k \Big) \Big( M(-+, \lambda \bar \lambda) - M(+-, \lambda \bar \lambda)\Big) \,.
\end{eqnarray}
Let us introduce the shorthand notation
\begin{eqnarray}
M^{\lambda \bar \lambda}_{ik} = \delta_{ik} M^{(0,+)}_{\lambda \bar \lambda} - i \epsilon_{ik} M^{(0,-)}_{\lambda \bar \lambda} + P^\parallel_{ik} M^{(2,+)}_{\lambda \bar \lambda} + i P^\perp_{ik} M^{(2,-)}_{\lambda \bar \lambda} \, .
\end{eqnarray}
Here
\begin{eqnarray}
\delta_{ik} &=& \hat x_i \hat x_k + \hat y_i \hat y_k \, , \, 
\epsilon_{ik} = \hat x_i \hat y_k - \hat y_i \hat x_k  \, ,
P^\parallel_{ik} = \hat x_i \hat x_k - \hat y_i \hat y_k \, , \, 
P^\perp_{ik} =  \hat x_i \hat y_k + \hat y_i \hat x_k  \,.
\end{eqnarray}
Then, in the cross-section of interest, we need
\begin{eqnarray}
\sum_{\lambda \bar \lambda} M_{ik}^{\lambda \bar \lambda} M_{jl}^{\lambda \bar \lambda \dagger} &=&
\delta_{ik} \delta_{jl} \sum_{\lambda \bar \lambda} \Big| M^{(0,+)}_{\lambda \bar \lambda} \Big|^2 +
\epsilon_{ik} \epsilon_{jl} \sum_{\lambda \bar \lambda} \Big| M^{(0,-)}_{\lambda \bar \lambda} \Big|^2 \nonumber \\
&+& P^\parallel_{ik}P^\parallel_{jl} \sum_{\lambda \bar \lambda} \Big| M^{(2,+)}_{\lambda \bar \lambda} \Big|^2
+  P^\perp_{ik}P^\perp_{jl} \sum_{\lambda \bar \lambda} \Big| M^{(2,-)}_{\lambda \bar \lambda} \Big|^2 \,.
\end{eqnarray}
Here we decomposed the $\gamma \gamma \to l^+ l^-$ amplitude into 
channels of total angular momentum projection $J_z = 0, \pm 2$ and 
even and odd parity.
The explicit expressions for the squares of amplitudes, in terms of cm-scattering angle $\theta$ read:
\begin{eqnarray}
\sum_{\lambda \bar \lambda} \Big| M^{(0,+)}_{\lambda \bar \lambda} \Big|^2 &=&  g^4_{\rm em} {8 (1-\beta^2) \beta^2 \over (1 - \beta^2 \cos^2 \theta)^2 }  \, , \, 
\nonumber \\
\sum_{\lambda \bar \lambda} \Big| M^{(0,-)}_{\lambda \bar \lambda} \Big|^2 &=&  g^4_{\rm em}
{ 8(1 - \beta^2) \over (1 - \beta^2 \cos^2 \theta)^2 } \, ,
\nonumber \\
\sum_{\lambda \bar \lambda} \Big| M^{(2,+)}_{\lambda \bar \lambda} \Big|^2 &=& g^4_{\rm em} { 8 \beta^2 \sin^2 \theta \over (1 - \beta^2 \cos^2 \theta)^2 } \Big( 1 - \beta^2 \sin^2 \theta \Big) \, , 
\nonumber \\
\sum_{\lambda \bar \lambda} \Big| M^{(2,-)}_{\lambda \bar \lambda} \Big|^2 &=& g^4_{\rm em} { 8 \beta^2 \sin^2 \theta \over (1 - \beta^2 \cos^2 \theta)^2 } \, ,
\end{eqnarray}
where $g^2_{\rm em} = 4 \pi \alpha_{\rm em}$, and
\begin{eqnarray}
\beta = \sqrt{1- {4m_l^2 \over M^2}} 
\end{eqnarray}
is the lepton velocity in the dilepton cms-frame. 
Notice that in the ultrarelativistic limit $\beta \to 1$, the $|J_z| =2$ terms dominate, while for
$\beta \ll 1$, relevant for heavy fermions, the
$J_z=0$ components are the leading ones.
A brief comment on the approach used in Ref.\cite{Klein:2020jom} is in order:
the helicity structure used for the hard 
$\gamma \gamma \to l^+ l^-$ process in \cite{Klein:2020jom} corresponds to the sum of our $|J_z|= 2$ terms for $\beta \to 1$.  

The phase space of the dileptons can be parametrized
by the rapidities $y_{1,2}$ and transverse momenta
$\bp_1, \bp_2$ of leptons.
Then the cross section fully differential in lepton variables is obtained as
\begin{eqnarray}
{d \sigma [{\cal C}]\over dy_1 dy_2 d^2\bp_1 d^2\bp_2} &=& 
\int {d^2 \bQ \over 2 \pi}  w(Q; b_{\rm max}, b_{\rm min}) 
 \int {d^2\bq_1 \over \pi} {d^2\bq_2 \over \pi} \, \delta^{(2)}(\bp_1 + \bp_2-\bq_1 - \bq_2)
\nonumber \\
&\times&   E_i \Big(\omega_1, \bq_1 + {\bQ \over 2} \Big) E^*_j \Big(\omega_1, \bq_1 - {\bQ \over 2} \Big)
E_k \Big(\omega_2, \bq_2 - {\bQ \over 2} \Big) E^*_l \Big(\omega_2, \bq_2 + {\bQ \over 2} \Big) \nonumber \\
&\times& 
{1 \over 16 \pi^2 \hat s^2}  \sum_{\lambda \bar \lambda} M^{\lambda \bar \lambda}_{ik} M^{\lambda \bar \lambda \dagger}_{jl} \, .
\end{eqnarray}
The delta-functions in the phase space element determine the photon energies as:
\begin{eqnarray}
\omega_1 &=& \frac{1}{2} \Big( m_{1t} \exp(+y_1) + m_{2t} \exp(+y_2) \Big) \, , 
\omega_2 = \frac{1}{2} \Big( m_{1t} \exp(-y_1) + m_{2t} \exp(-y_2) \Big) \, .
\label{omegas}
\end{eqnarray}
where $m_{it} = \sqrt{\bp_i^2 + m_l^2}$.
We perform the multidimensional integration using the VEGAS Monte Carlo
method \cite{Lepage:1977sw}.

\section{Results of the new approach versus 
experimental data}


In this section we present predictions for the centrality dependence of $l^+ l^-$ production
Au-Au collisions at RHIC energy ($\sqrt{s_{NN}}$=200 GeV)
and 
Pb-Pb collisions at LHC energy ($\sqrt{s_{NN}}$=5.02 TeV). 
The results of our approach will be compared to STAR, ALICE and ATLAS experimental data.

\begin{figure}[!h] 
(a)\includegraphics[scale=0.35]{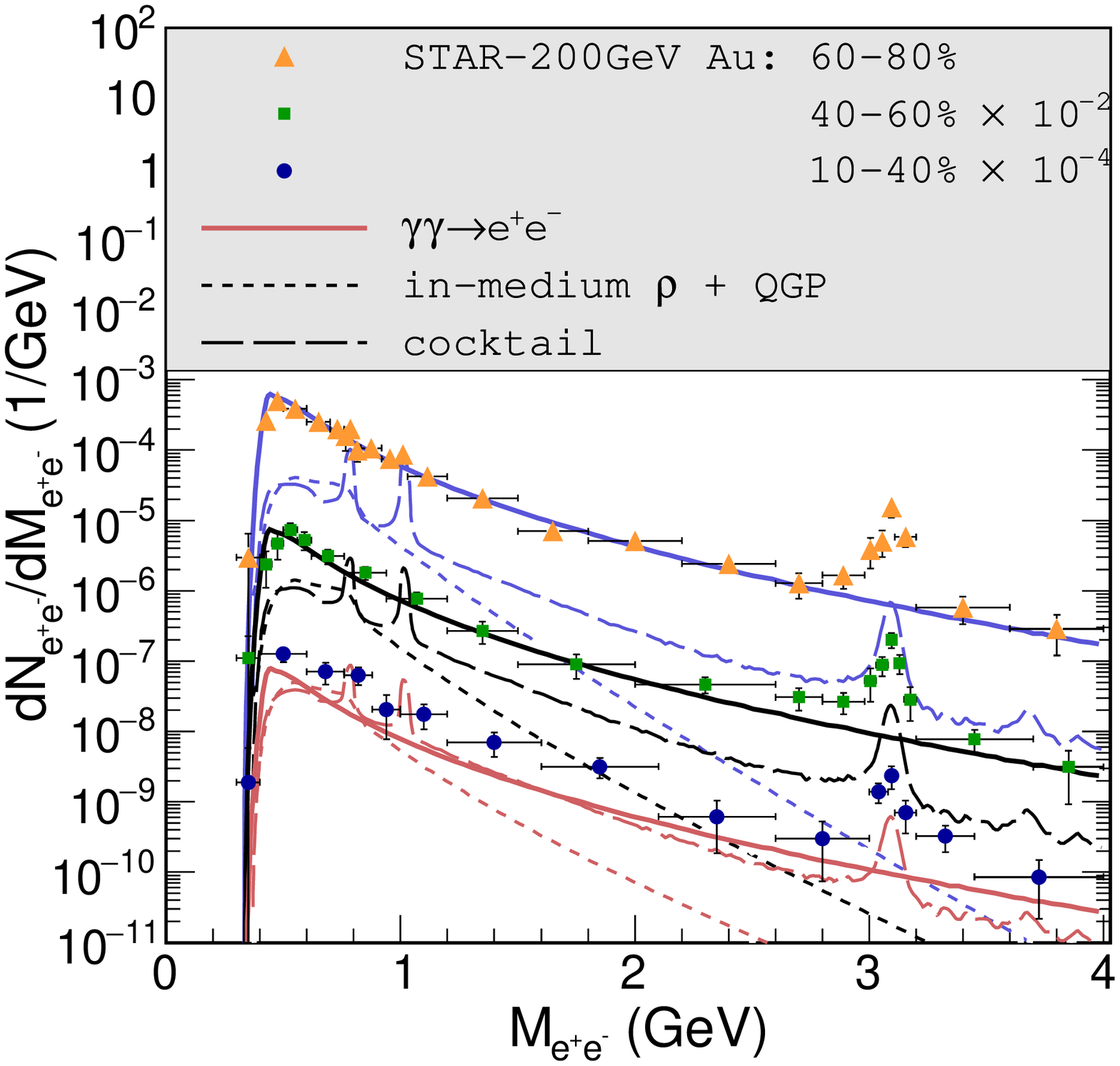}
(b)\includegraphics[scale=0.35]{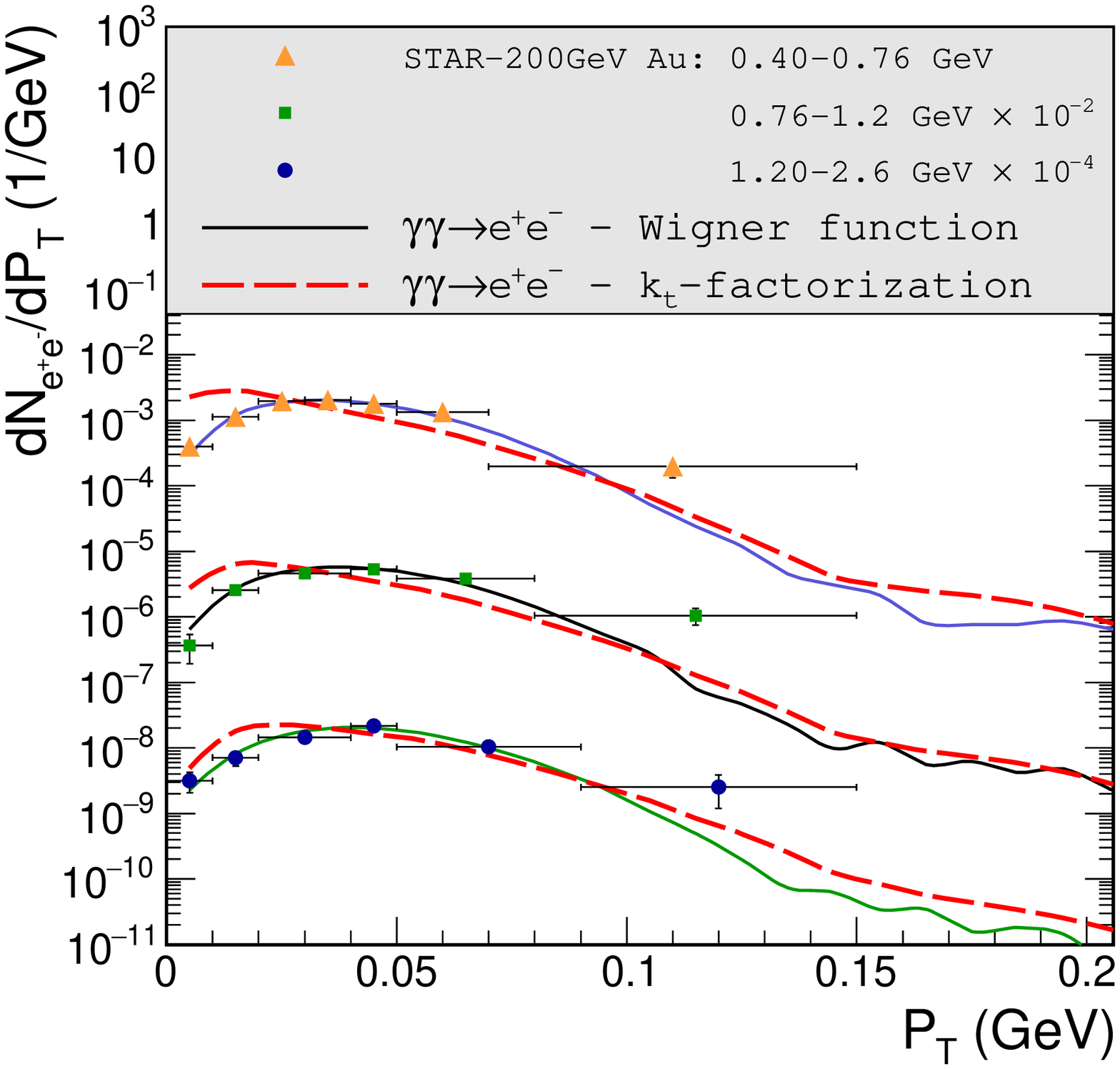}
\caption{
Theoretical predictions vs. STAR experimental data at RHIC energy ($\sqrt{s}$ = 200 GeV) \cite{Adam:2018tdm}. (a) Dielectron invariant mass spectra for three ranges of centrality: 60 - 80 \% 
(upper curve), 40 - 60 \% (middle curve) and 10 - 40 \% (lower curve)  \cite{Adam:2018tdm}.
Thermal radiation (dotted lines) and hadronic cocktail (dashed lines) contributions are compared with $\gamma \gamma \to e^+ e^-$ process \cite{Klusek-Gawenda:2018zfz}. 
(b) Distribution of transverse momentum of dielectron pair
for three invariant mass limits \cite{Adam:2018tdm}.
$k_T$-factorization result (dashed lines) \cite{Klusek-Gawenda:2018zfz} is  shown for comparison to the new  results (solid lines). The centrality is in the limit: (60-80)$\%$}
\label{fig:STAR}
\end{figure}

We start the presentation of our results from the invariant mass distribution
of low-$P_T$ dielectrons.
In Fig.\ref{fig:STAR}(a) we show our $\gamma\gamma$-fusion
results (solid lines) together with
the STAR experimental data \cite{Adam:2018tdm} for three different ranges of
centrality as defined in the STAR experiment.
The calculation of the invariant mass distribution reproduces very well the results of our earlier work \cite{Klein:2018fmp}.
The photon-photon process dominates for peripheral collisions. 
For the most central
collisions, (10 - 40 )\%, the $\gamma \gamma$ process starts to underestimate  the experimental data.
The detailed discussion of thermal radiation and hadronic cocktail is found in \cite{Klusek-Gawenda:2018zfz} and
must not be repeated here. 
We stress that the theoretical results are calculated including STAR experimental cuts i.e. $p_{t,e}>0.2$~GeV, $|\eta_e|<1$, $|y_{e^+e^-}|<1$ and $P_T<0.15$~GeV. 

In Fig.\ref{fig:STAR}(b) we show the distribution in transverse
momentum of the dielectron pairs for three centrality intervals.
We get very good description of the low-$P_T$ enhancement, 
within the newly presented approach.

In our previous work \cite{Klusek-Gawenda:2018zfz} we calculated the
$P_T$-distribution in what we dub a $k_T$-factorization approach,
where the transverse momentum of dileptons involves a
convolution of the $\bq$-dependent photon fluxes, schematically:
\begin{eqnarray}
N_{l^+ l^-}(\bP) \propto \int {d \omega_1 \over \omega_1} {d \omega_2 \over \omega_2} \int {d^2 \bq_1 \over \pi} {d^2 \bq_2 \over \pi}
\delta^{(2)}(\bP - \bq_1 -\bq_2) N(\omega_1, \bq_1) N(\omega_2, \bq_2) \sigma_{ \gamma \gamma \to l^+ l^-}(4 \omega_1 \omega_2) \, . \nonumber \\
\label{eq:kt-fact}
\end{eqnarray}

While our previous calculations in
\cite{Klusek-Gawenda:2018zfz} convincingly demonstrated the dominance of the $\gamma \gamma$ process at low $P_T$, the peak of the distribution obtained from Eq.~(\ref{eq:kt-fact}) is systematically at too low values of $P_T$, and neglects the correlation of $P_T$ with centrality.

Our new theoretical results give an excellent description of both the shape
and normalization of the low-$P_T$ enhancement.
In our approach here we use a charge form factor of the nucleus obtained from a realistic charge distribution described in \cite{KlusekGawenda:2010kx}.
The form factor almost coincides with the one used in \textsc{STARlight} simulation code~\cite{Klein:2016yzr}.  

\begin{figure}[!h]
\includegraphics[scale=0.5]{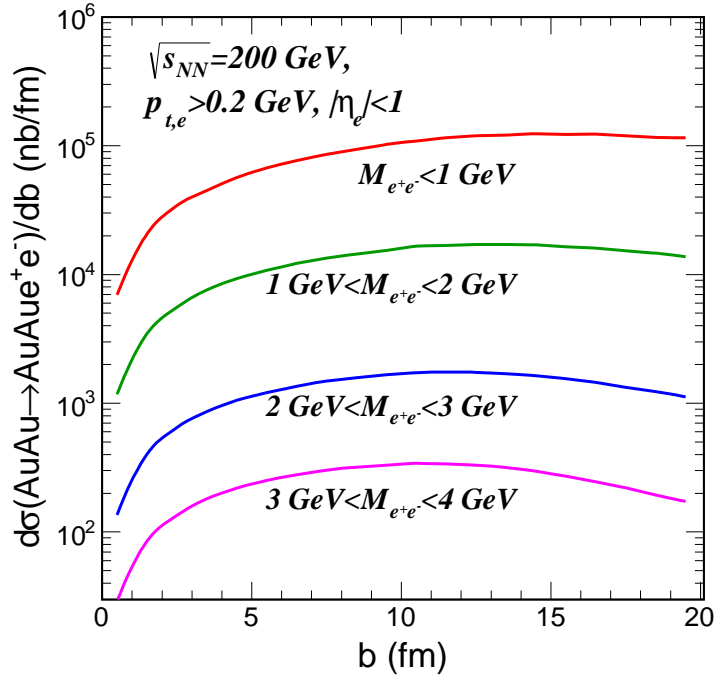}
\caption{Impact parameter dependence for $\sqrt{s_{NN}}$ = 200 GeV
and different windows of dielectron invariant mass.}
\label{fig:dsig_db_STAR}
\end{figure}

For completeness, 
in Fig.\ref{fig:dsig_db_STAR} we show cross section for the $Au Au \to e^+ e^- Au Au$ process as a function of impact parameter. Here we have taken kinematical cuts
adequate for the STAR experiment.
We note that the shape in impact parameter depends on the window of dielectron mass. 
The impact parameter cannot be directly measured. 
As previously done in \cite{Klusek-Gawenda:2018zfz}, we use an optical Glauber model \cite{Miller:2007ri} to estimate geometric quantities. 
For normalization we need the total hadronic inelastic cross section. We use the following values of cross sections: $\sigma_{AuAu}(\sqrt{s_{NN}}=200~\mbox{GeV}) = 6\, 936$~mb and $\sigma_{PbPb}(\sqrt{s_{NN}}=5.02~\mbox{TeV}) = 7\, 642$~mb.

\begin{figure}[!h]
\includegraphics[scale=0.5]{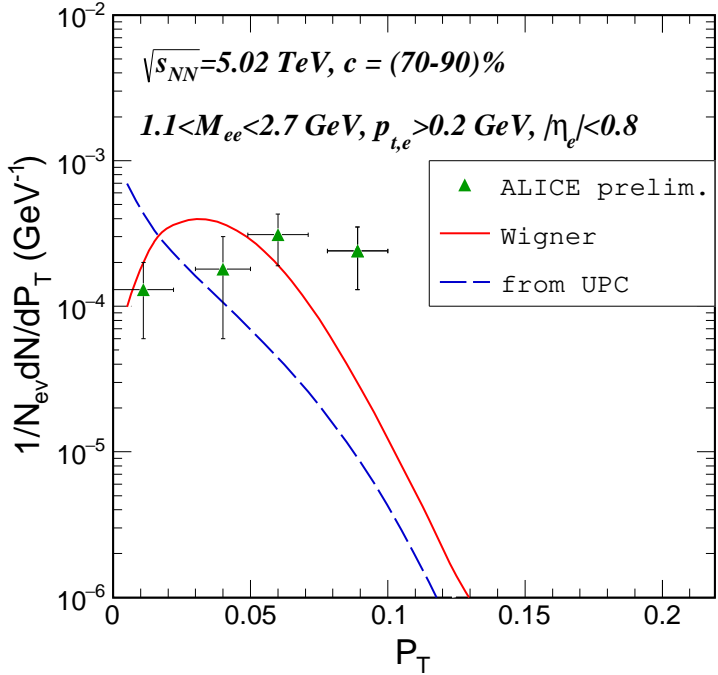}
\caption{Distribution in transverse momentum of the dielectron pair
for $\sqrt{s}$ = 5.02 TeV. The results of the present approach
is shown by the solid line.
The preliminary ALICE data \cite{Lehner:2019amb} are shown for comparison.}
\label{fig:dsig_dptsum_ALICE}
\end{figure}

We now proceed to LHC energies.
In Fig.~\ref{fig:dsig_dptsum_ALICE} we show our results compared to the preliminary experimental results obtained by the ALICE collaboration \cite{Lehner:2019amb}. 
We get a similarly good description of the preliminary ALICE data as for the case of the STAR data.
For illustration we show also result of the $k_T$-factorization (dashed line)
proposed in \cite{Klusek-Gawenda:2018zfz}, which clearly fails to describe the position of the peak. 
Indeed, the peak of the $P_T$ distribution predicted by the $k_T$-factorization formula Eq.~(\ref{eq:kt-fact}) runs away towards smaller and smaller $P_T$ with increasing energy.
This is related to the fact, that in the photon distribution
\begin{eqnarray}
N(\omega,\bq) \propto |\bE (\omega,\bq)|^2 \propto {\bq^2 \over [\bq^2 + (\omega/\gamma)^2 ]^2} F^2_{\rm{ch}}(\bq^2 + (\omega/\gamma)^2 ) \, ,
\end{eqnarray}
the ``cutoff'' $\omega/\gamma$ decreases with increasing cm-energy (or boost $\gamma$ of the ion).

Consequently, for the much lower STAR energies the $k_T$-factorization approach is only
slightly worse than that in the new approach (see Fig.~\ref{fig:STAR}(b)).

\begin{figure}
\includegraphics[scale=0.55]{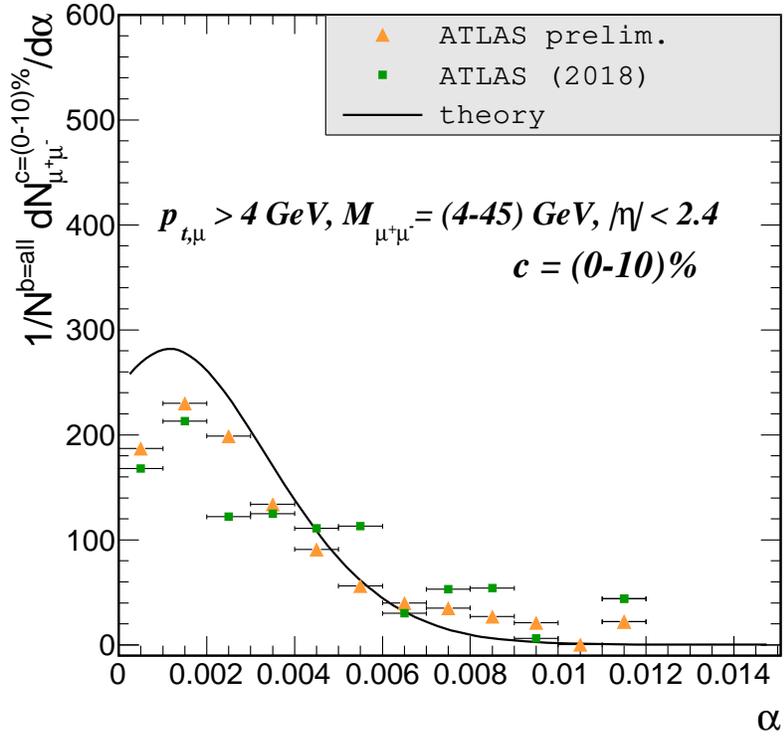}
\caption{Acoplanarity distribution for $\mu^+ \mu^-$ production 
for the kinematics of the ATLAS experiment. For comparison we show preliminary results \cite{ATLAS:2019vxg} as well as previously published ones \cite{Aaboud:2018eph}.}
\label{fig:dsig_dAco_twoATLAS}
\end{figure}

Finally in Fig.\ref{fig:dsig_dAco_twoATLAS} we show the acoplanarity distribution
for dimuon production as measured by the ATLAS collaboration \cite{Aaboud:2018eph,ATLAS:2019vxg}. 
The acoplanarity is defined as:
\begin{equation}
\alpha = 1-{|\Delta \phi_{l^+ l^-}| \over \pi}  \; ,
\end{equation}
where $\Delta \phi_{l^+ l^-}$ is the difference of the azimuthal angles of two leptons. 
The measurement of the dimuon production via $\gamma\gamma$ scattering process was done at rather large transverse momenta ($p_{t,\mu}>4$~GeV) of leptons. Our previous study of semi-central collisions at the LHC \cite{Klusek-Gawenda:2018zfz} has shown that the dilepton invariant mass (up to $1.5$~GeV) thermal radiation occurs only at centrality smaller than 50~\%.
Consequently, we  obtain a successful description of data by $\gamma\gamma$-fusion alone for the high invariant mass even for low centrality.
Our result is also in surprisingly good agreement for the case of very central collisions  ($c=(0-10)\%$). 
Here we get better agreement with the new preliminary data. 

\begin{figure}[!h]
\includegraphics[scale=0.265]{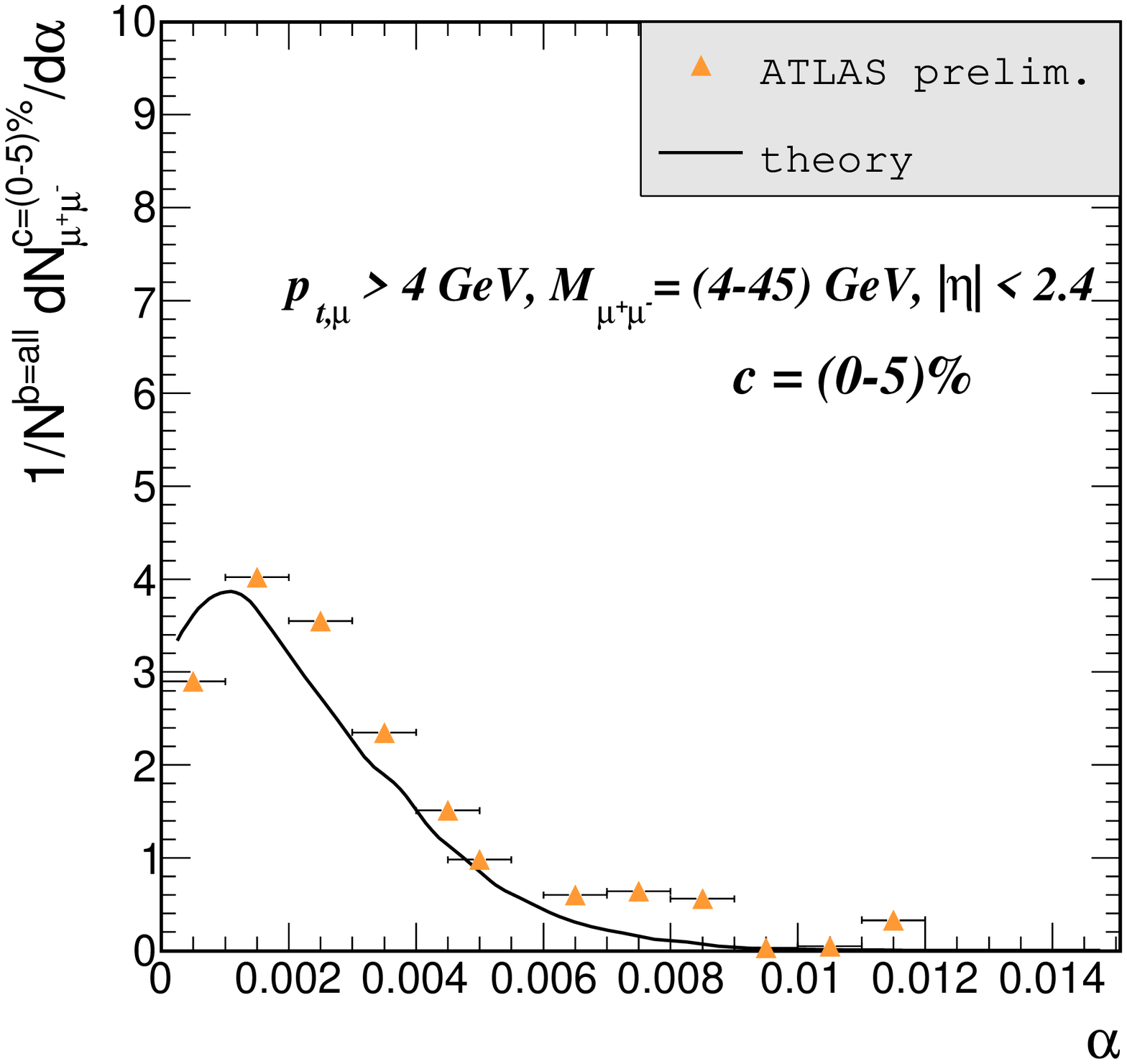}
\includegraphics[scale=0.265]{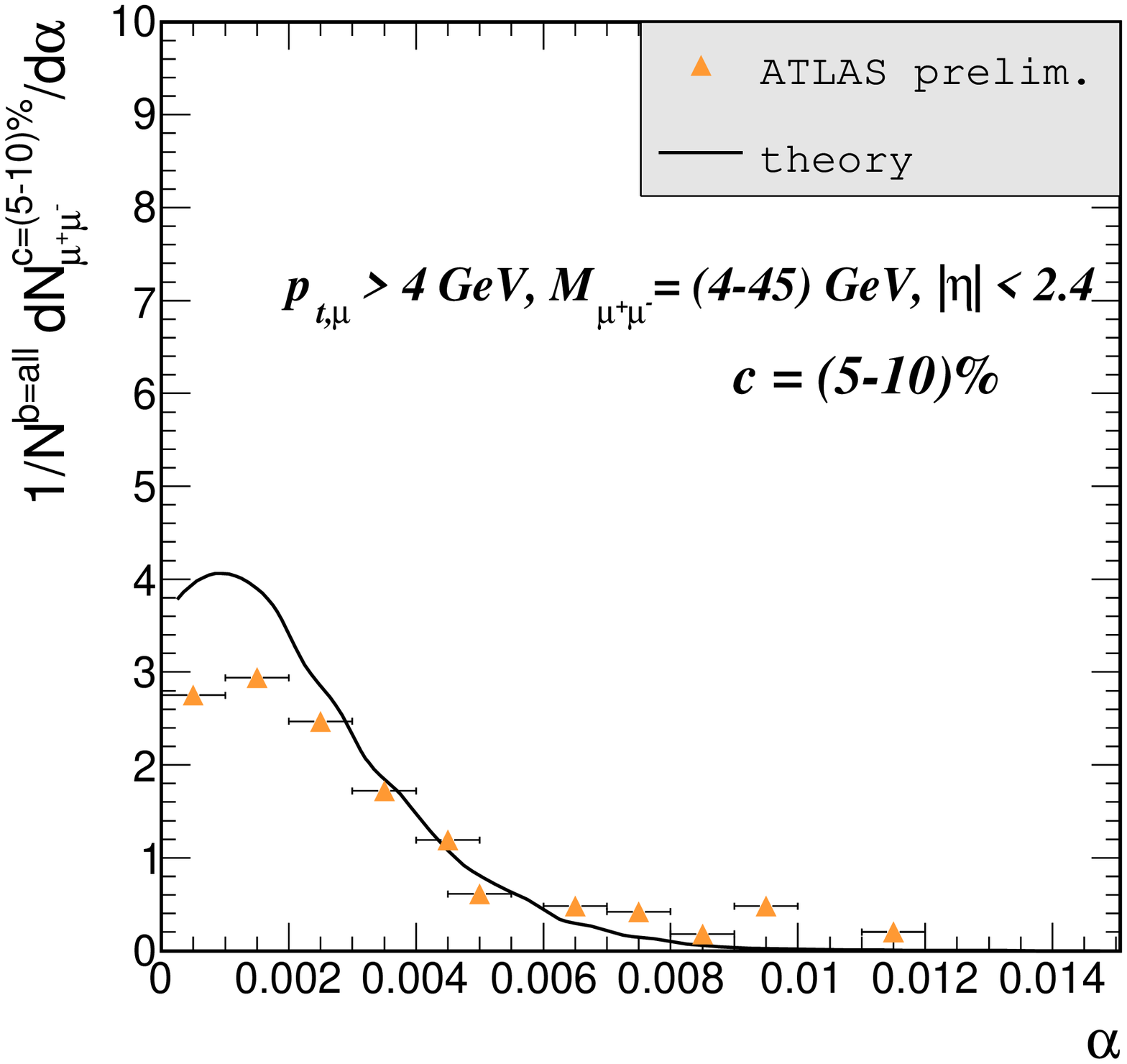}
\includegraphics[scale=0.265]{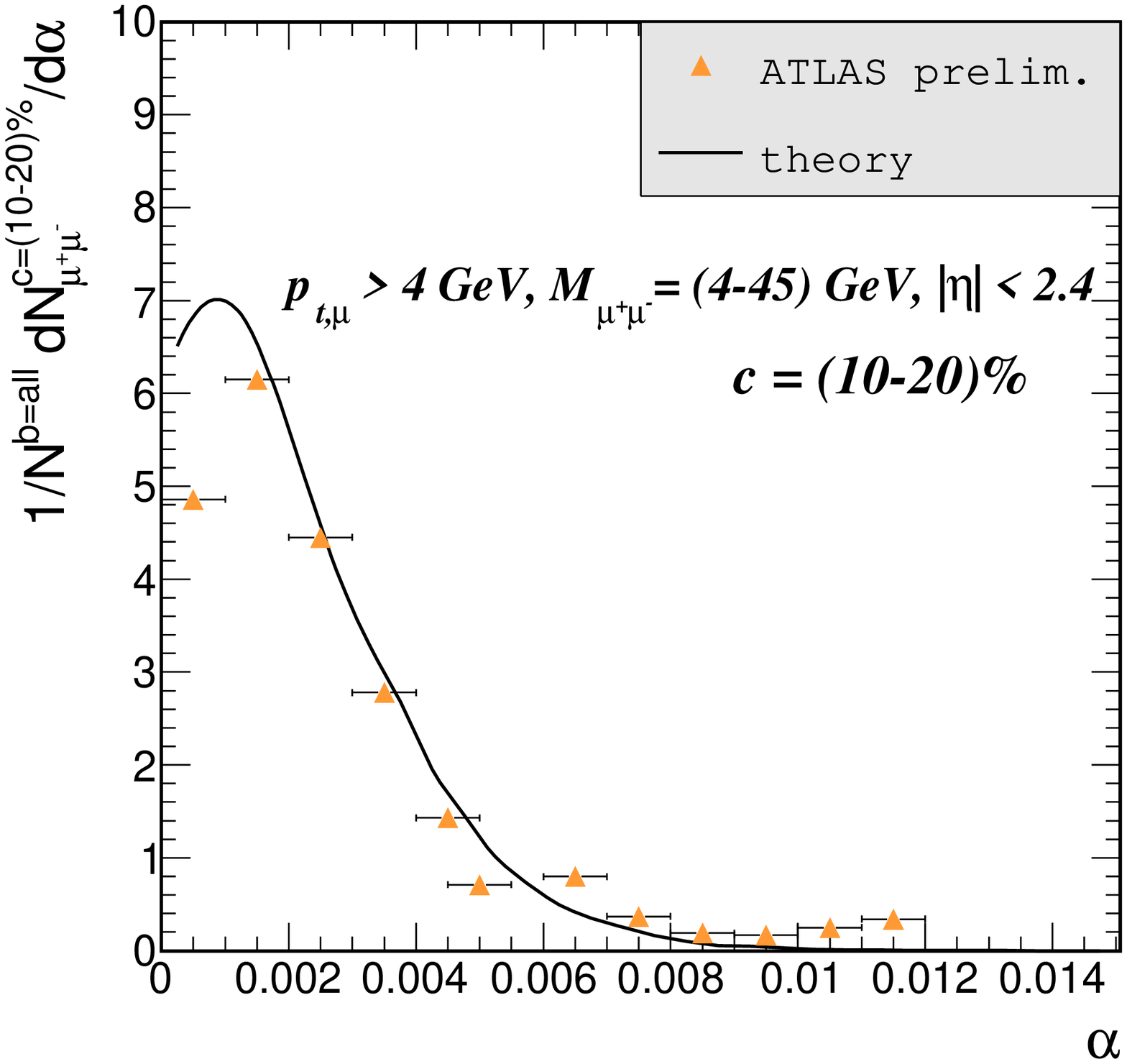}\\
\includegraphics[scale=0.265]{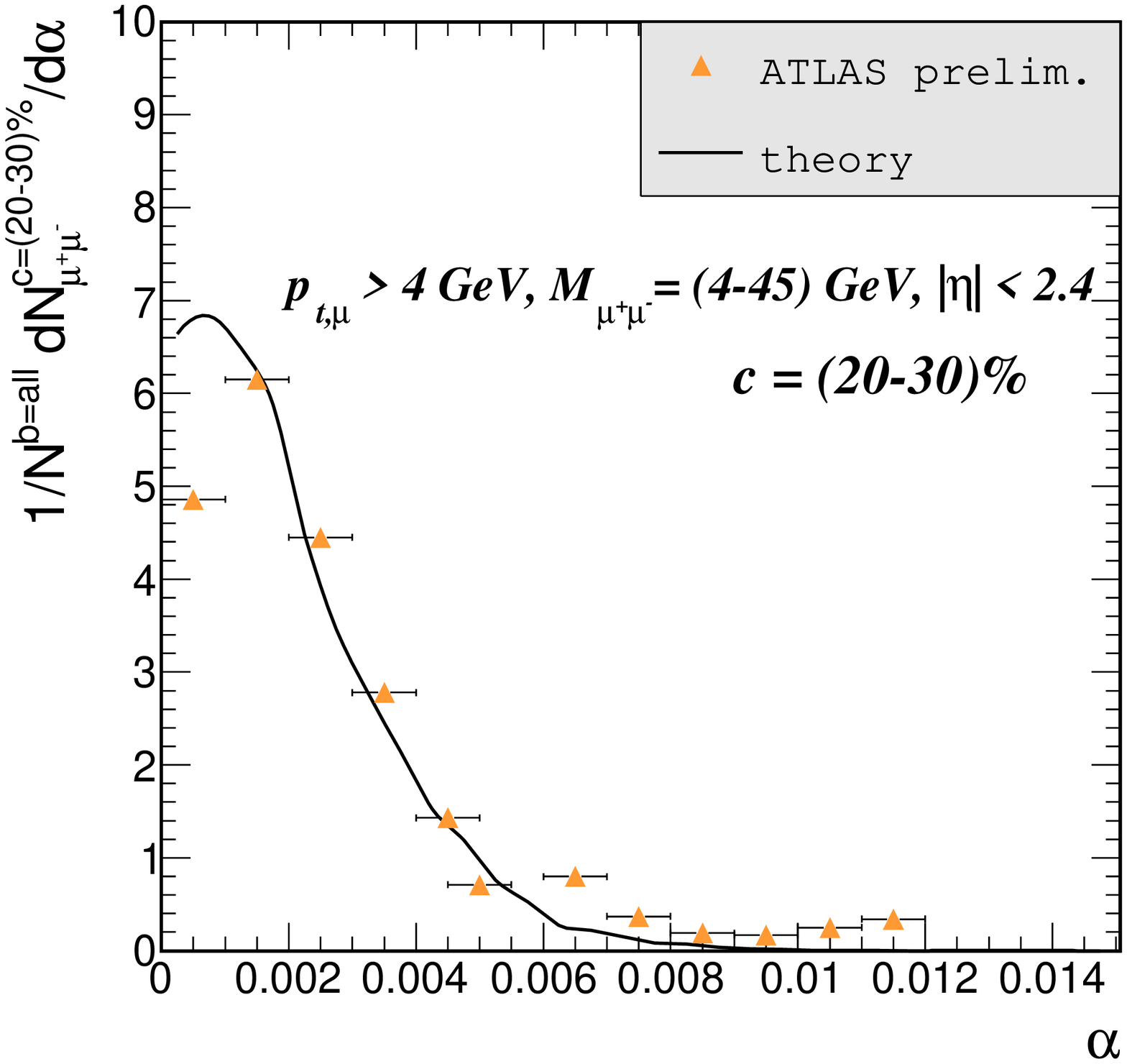}
\includegraphics[scale=0.265]{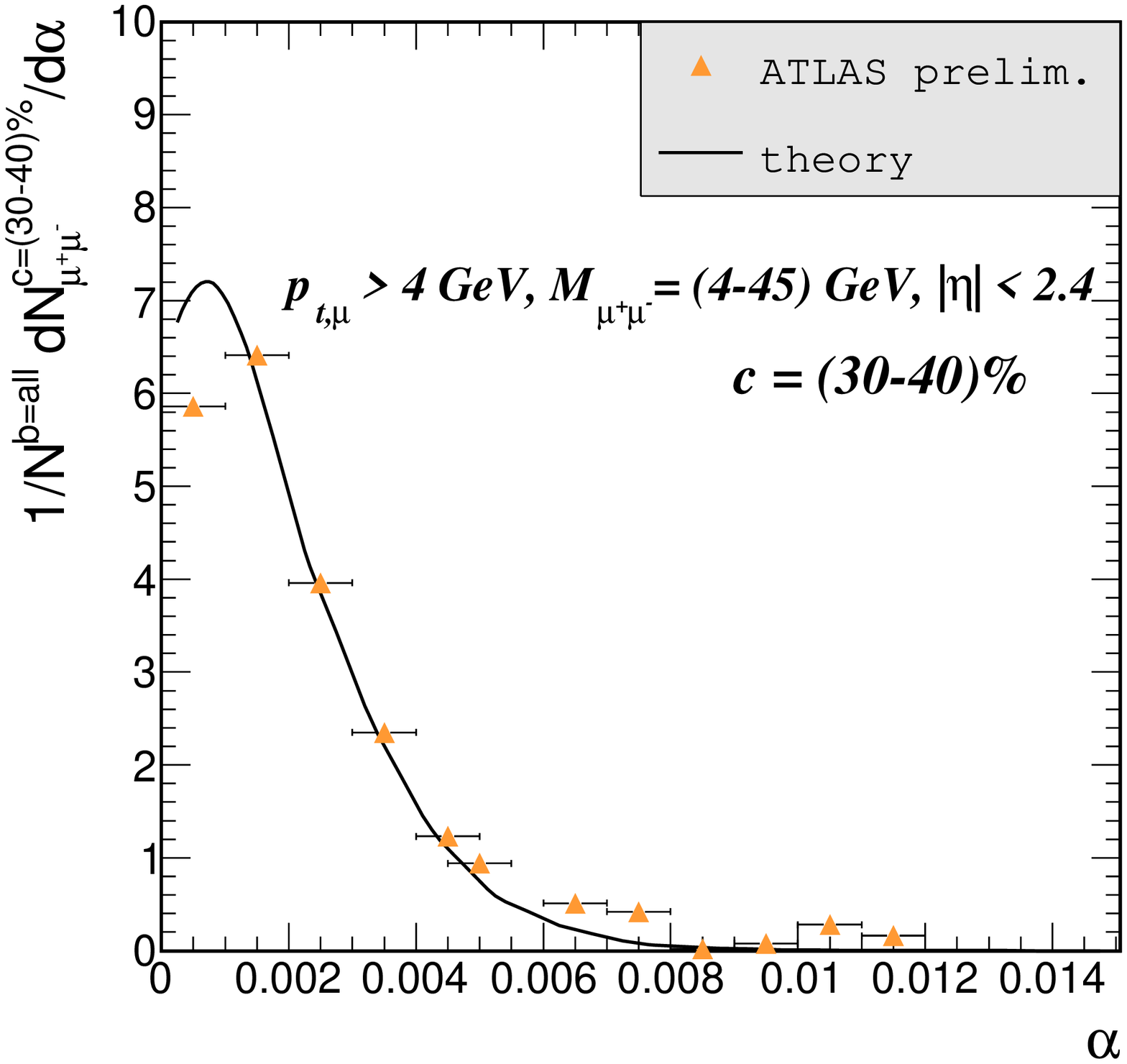}
\includegraphics[scale=0.265]{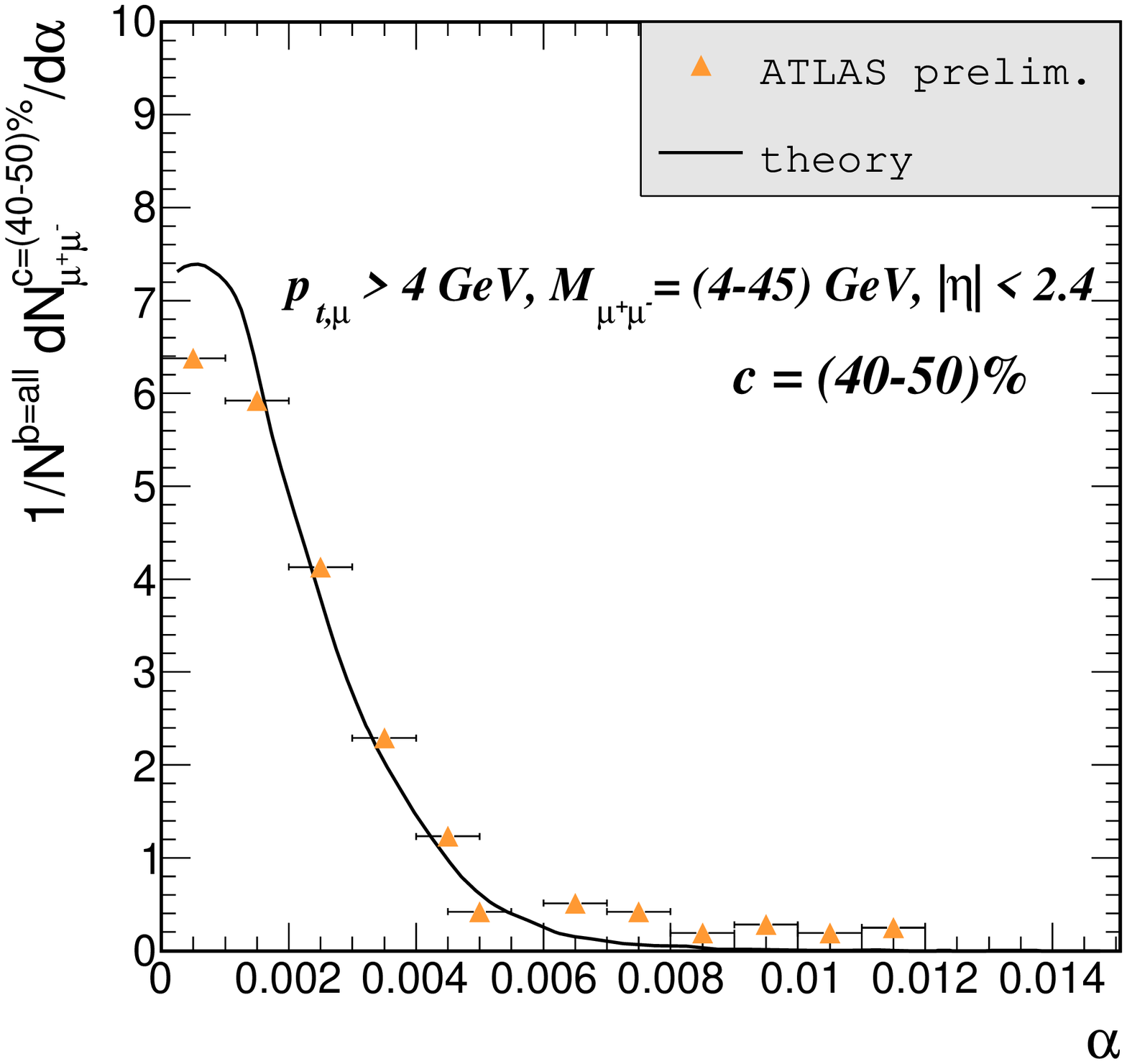}\\
\includegraphics[scale=0.265]{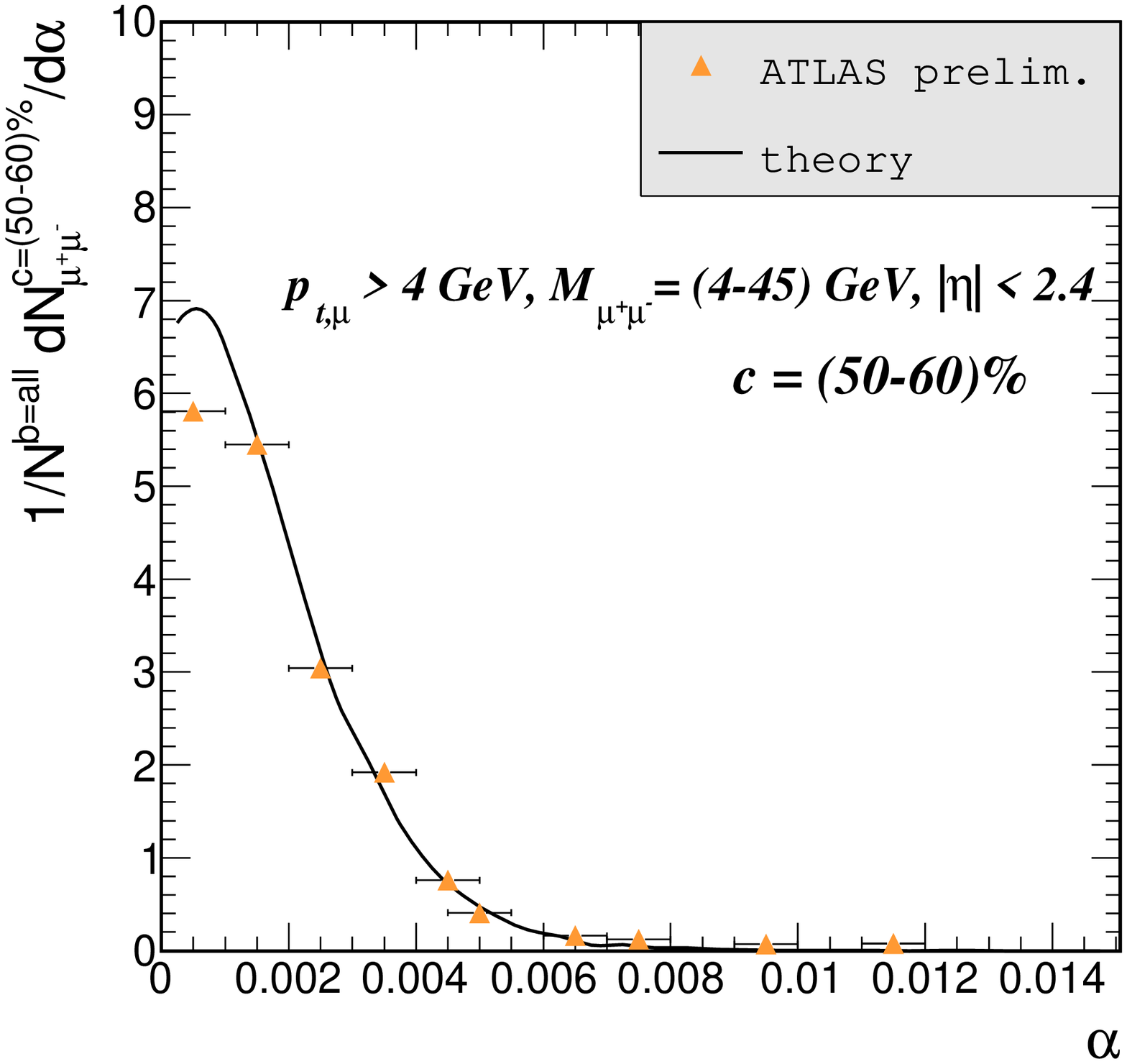}
\includegraphics[scale=0.265]{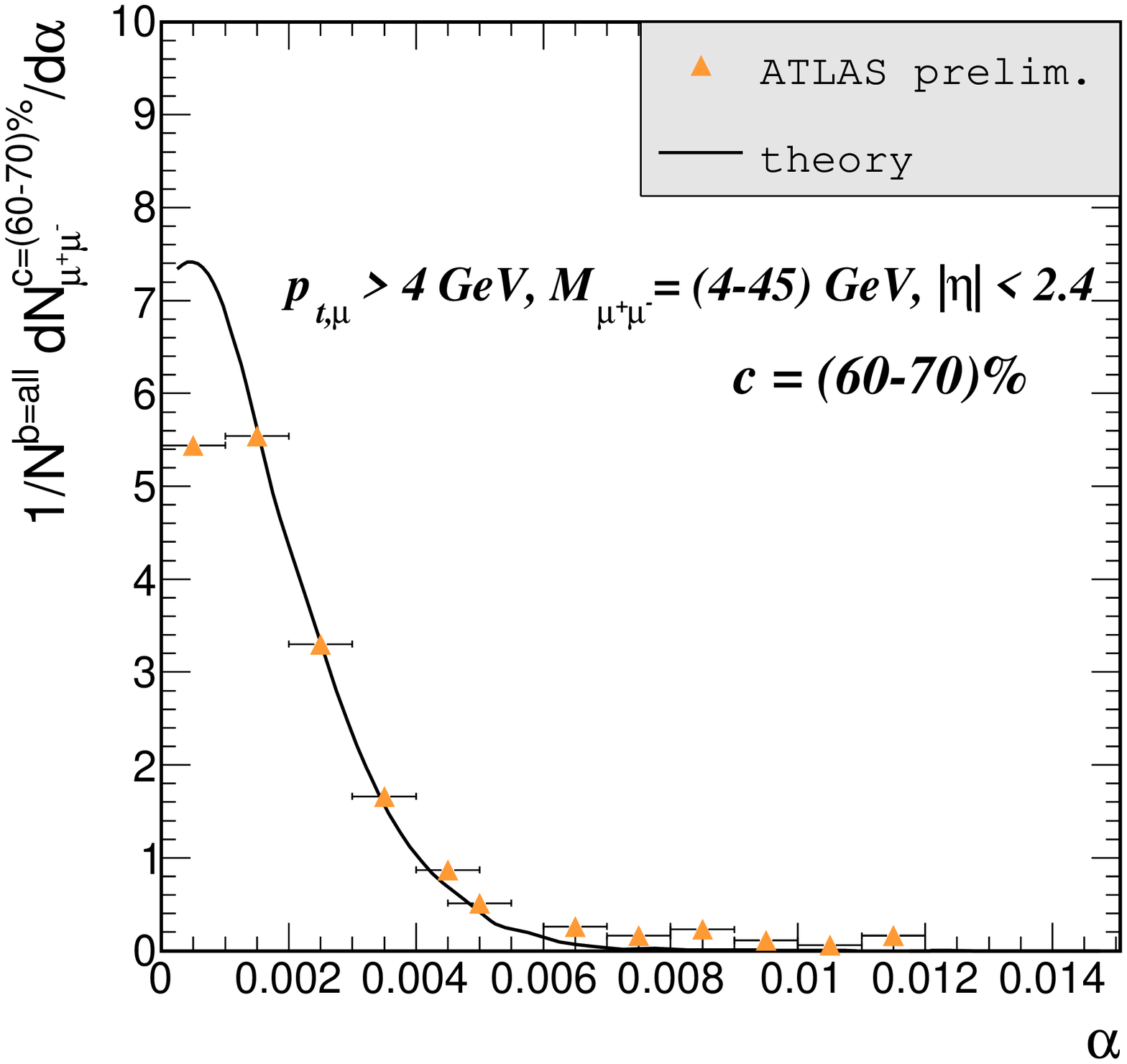}
\caption{Acoplanarity distribution for $\mu^+ \mu^-$ production 
for the kinematics of the ATLAS experiment 
\cite{Aaboud:2018eph, ATLAS:2019vxg}. The solid line represents result of the approach presented
in this paper. Each panel corresponds to a different centrality class.}
\label{fig:dsig_dAco_ATLAS}
\end{figure}

In Fig.~\ref{fig:dsig_dAco_ATLAS} we show the results for nine different centrality intervals, from central up to peripheral heavy-ion collision. We have got a very nice description of the data including correct normalization and shape of the distributions.
At larger acoplanarity there is a long tail possibly due to soft photon emission \cite{Klein:2020jom}.
This tail cannot be, however, seen in the linear scale used here and is not included in our calculation.

\section{Conclusion}

In this paper we have presented a formalism how to calculate differential distributions of leptons produced
in semi-central ($b < 2 R_A$) nucleus-nucleus collisions for a given centrality.
In this approach the differential cross section is calculated using the complete polarization density matrix of photons resulting from the Wigner distribution formalism.

We have presented results of calculation of several differential
distributions such as
invariant mass of dileptons, dilepton transverse momentum and acoplanarity for different
regions of centrality.
The results of the calculations have been compared to experimental
data of the STAR, ALICE and ATLAS collaboration.
A good agreement has been achieved in all cases.
Our approach gives much better agreement with experimental data
than the previous approaches used in the literature. 
Recently the CMS collaboration measured modifications of $\alpha$ distributions \cite{Sirunyan:2020vvm} 
correlated with neutron multiplicity. 
A very new ATLAS study also presents the dimuon cross
sections in the presence of forward and/or backward neutron production \cite{Aad:2020dur}.
This goes beyond present studies, and we plan such studies it in the future. Our formalism can be readily extended to such processes, using the impact parameter distributions obtained in \cite{Klusek-Gawenda:2013ema}.

We have obtained a good description of the data without introducing additional
final state rescattering of leptons in the quark-gluon plasma.
More work is necessary to identify observables that can probe electromagnetic properties of the QGP.

\section*{Acknowledgements}
This work was partially supported by the Polish National Science Center
under grant No. 2018/31/B/ST2/03537 and by the Center for Innovation and Transfer of Natural Sciences and Engineering
Knowledge in Rzesz\'ow (Poland).
\bibliographystyle{unsrturl}
\bibliography{biblio}

\end{document}